\documentclass[
  aip,
  chaos,
  reprint,
]{revtex4-1}

\usepackage{graphicx}
\usepackage{dcolumn}
\usepackage{bm}

\usepackage{physics}
\usepackage[T1]{fontenc}
\usepackage[utf8]{inputenc} 
\usepackage{mathptmx}
\usepackage{amsmath} 
\usepackage{amssymb}
\usepackage{etoolbox}
\usepackage{hyperref}
\usepackage{amsfonts}

\makeatletter
\def\@email#1#2{%
 \endgroup
 \patchcmd{\titleblock@produce}
  {\frontmatter@RRAPformat}
  {\frontmatter@RRAPformat{\produce@RRAP{*#1\href{mailto:#2}{#2}}}\frontmatter@RRAPformat}
  {}{}
}%
\makeatother
\begin{document}
\preprint{AIP}

\title[Ergodic dynamics in iterated quantum protocols]{Ergodic dynamics in iterated quantum protocols}

\author{Attila Portik}
\affiliation{HUN-REN Wigner Research Centre for Physics, 1525 P.O. Box 49, Hungary}
\affiliation{E\"otv\"os Lor\'and University, P.O.\ Box 32, H-1518, Hungary}
\affiliation{Qutility @ Faulhorn Labs, H-1117, Hungary}
\author{Orsolya Kálmán}
\affiliation{HUN-REN Wigner Research Centre for Physics, 1525 P.O. Box 49, Hungary}
\author{Tamás Kiss}
\affiliation{HUN-REN Wigner Research Centre for Physics, 1525 P.O. Box 49, Hungary}

\date{\today}

\begin{abstract}
We study measurement-induced nonlinear dynamics generated by an iterated quantum protocol combining an entangling gate, a single-qubit rotation, and post-selection. For pure single-qubit inputs, a particular choice of the single-qubit unitary yields globally chaotic, strongly mixing dynamics that explores the entire Bloch sphere, providing a physical realization of ergodic behavior in a complex map. We extend the analysis to realistic, noisy preparation by considering mixed initial states and the induced nonlinear evolution inside the Bloch sphere. Numerical results show that the maximally mixed state is an attractor for mixed inputs, although many trajectories exhibit transient increases in purity before ultimately converging. To quantify robustness against noise, we introduce a practical notion of quasi-ergodicity: ensembles prepared in a small angular patch at fixed purity rapidly spread to cover all directions, while the purity gradually decreases toward its minimal value. By varying the final single-qubit gate, we identify a broad family of protocols that remain ergodic-like for pure states, supported by consistent diagnostics including the absence of attracting cycles, agreement of time and ensemble statistics, rapid spreading from localized regions, and exponential sensitivity to initial conditions. Away from the special globally mixing case, the mixed-state dynamics can change qualitatively: for most ergodic-like parameters, a finite subset of noisy inputs is driven toward purification rather than complete mixing, demonstrating the coexistence of statistical mixing and purification within a single iterated protocol.
\end{abstract}

\maketitle

\begin{quotation}
Iterated maps, chaos, and ergodicity are central elements of nonlinear dynamics, and iterated quantum protocols provide a concrete physical approach to realizing such phenomena through measurement-induced nonlinearity. The combination of an entangling quantum gate and post-selection, applied to identical input states, leads to an essentially nonlinear protocol. A controlled not gate combined with a specific single-qubit gate is known to exhibit ergodic dynamics for pure single-qubit inputs. All initial states lead to chaotic behavior, which is ergodic in the following sense: initial states from a small neighborhood of any given initial state will fully cover the state space after a critical number of iterative steps. Ergodicity is a consequence of the fact that the dynamics can be represented by a Latt\`es map. For these maps, the forward orbit of a single initial point is everywhere dense in the state space, thus satisfying the usual criterion of ergodicity, the equivalence of time- and ensemble-averages. We study the time-evolution of mixed initial states, containing some amount of incoherent noise. The completely mixed state is an attractive fixed point of the dynamics, and our numerical simulations convincingly suggest that the open set of mixed states is the basin of attraction. Nevertheless, applying a modified definition of ergodicity by choosing initial states from a small neighborhood with the same purity, we provide evidence that the dynamics is quasi-ergodic. In other words, small noise, which is always present in a practical realization of the protocol, will practically not destroy ergodicity in the experimentally relevant, finite-time sense. We examine the choice of the single-qubit gate and identify an extended parameter regime for which the above conditions, as well as the condition that the Lyapunov exponent is positive for all initial states, are fulfilled. We also show that the evolution for mixed initial states can be radically different for the Latt\`es map and for other ergodic parameters: we find numerical evidence that in the latter case, some mixed initial states will be purified. Together, these results demonstrate that iterated quantum protocols provide a physically accessible realization of canonical complex dynamical systems, with a tunable balance between chaotic mixing on the Bloch sphere and noise-dependent purification in the Bloch ball.
\end{quotation}

\section{Introduction}
\label{Introduction}

Ergodicity is a key concept in modelling the behavior of physical systems ~\cite{Moore}. It is closely related to predictability  of the evolution and, in turn, is at the heart of statistical physics. The fundamental idea behind the concept of ergodic evolution, first proposed by Boltzmann \cite{Boltzmann}, is that the state of the system eventually reaches all parts of the phase space \cite{Landau}. Another usual formulation of the concept of ergodicity requires that ensemble averages coincide with time averages. Under a rigorous mathematical formulation of ergodicity, a number of models have been proved to be ergodic \cite{Halmos,Viana}.

In quantum mechanics, the evolution of closed systems is governed by unitary dynamics, where the overlap of two quantum states will not change during the evolution, thereby excluding ergodicity.  Nevertheless, the concept of quantum ergodicity is applied in the literature, referring to semi-classical behavior of the eigenfunctions of the Schrödinger equation  ~\cite{chaosbook}. Random perturbations in open quantum systems can also lead to ergodic evolution ~\cite{filippov2017divisibility,grimaudo2020two,luchnikov2017quantum,karpat2021synchronization}.

The emergence of quantum information theory and especially the realization of the first noisy quantum computers opened another contextual framework to study more general types of quantum evolution ~\cite{Lloyd2000}. In this context, it is natural to make measurements on a part of the system (qubits) and then, according to the measurement result, let the evolution continue or abandon it (feed-forward in quantum computers). A typical example is entanglement distillation, where multiple copies of a system (e.g. pairs of qubits) in the same quantum state are processed and, depending on measurement results, some of them are kept while others are discarded ~\cite{Bennett1996, Bennett1996b, fang2019non}. This process may be repeated on the kept outcomes, leading to iterative schemes ~\cite{dur2007entanglement, guan2013reexamination}. The evolution of this type of dynamics is inherently nonlinear and may lead to truly chaotic behavior in the phase space of the system ~\cite{Schuster, abrams1998nonlinear}. Entanglement itself may evolve chaotically in such a scheme~\cite{Kiss2011}.

The above described iterative dynamics can be analyzed for single-qubit systems as well~\cite{BechmannP,Alber2001}. The evolution of pure states can conveniently be represented by iterations of a complex function~\cite{Kiss2006,kiss2006complex_Acta}. The most general form of the evolution, corresponding to a specific (entangling) core unitary, is represented by a quadratic rational function ~\cite{Gilyen2016}. A rich variety of dynamical behavior emerges when iterating a quadratic rational complex function, including chaotic evolution. For a given core unitary, the initial states of the qubit (represented by a point on the Bloch sphere) belong to one of two sets: the Fatou set -- roughly speaking -- contains all fixed points and cycles and their respective basins of attraction, while initial states in the Julia set evolve in an unstable manner and are considered chaotic in some sense. The Julia set is either of zero measure on the Bloch sphere or it is the whole sphere, containing all possible initial pure states. The latter case is usually considered ergodic in the mathematical literature ~\cite{Milnorbook}. A special type of ergodic maps is the Latt\`es map with stronger ergodic properties ~\cite{Lattes}. The zero measure Julia sets can form fractals, with a fractal dimension between $0$ and $2$ ~\cite{DetChaos}. 

The same protocols can be analyzed also for mixed initial quantum states, which is a natural assumption in a physical realization. Mixed initial states extend the geometrical representation to the interior of the Bloch sphere. A mixed initial state may be purified during the evolution, the same process for more qubits is related to entanglement distillation. In special cases, the fractal Julia sets can be extended to mixed initial states via defining a quasi-Julia set as either border points of convergence regions for fixed purity spheres or as the pre-images of initial states close to the Julia set. In these cases, the fractal suddenly disappears at a critical purity akin to a phase transition. The phase transition phenomenon is present in more general, higher order protocols as well, where more qubits are processed in one step \cite{Malachov_2019, PORTIK2022127999}.

In this paper, we analyze the behavior of ergodic protocols for mixed initial states. We find that in the Latt\`es case, all mixed initial states converge to the completely mixed state, in accordance with our earlier study~\cite{kalman2018sensitivity}. We show that this convergence is not monotonic with respect to the purity of the state; some mixed initial states first become slightly purified during the evolution before becoming completely mixed.  We study perturbations of the parameter in the Latt\`es scheme and find that it can lead to protocols which are also ergodic, even though they are not of the Latt\`es type in a strict sense. We numerically identify the parameter regime in the perturbed scheme where ergodicity is expected, by applying a variety of criteria: (i) the convergence of the critical points, (ii) the equivalence of ensemble and time averages, (iii) the spreading of initially close states over the phase space and (iv) the value of the Lyapunov exponent. Our simulations suggest that for non-Latt\`es type ergodic protocols there is always a set of mixed initial states, which do not converge to the completely mixed state, instead, they tend to get completely purified.

The paper is organized as follows. Section~\ref{Introduction} introduces measurement-induced nonlinear dynamics in iterated quantum protocols, reviews the ergodic pure-state case, and formulates the main questions for mixed initial states. In Section~\ref{Pure state dynamics in a Latt\`es-type ergodic quantum protocol} we discuss the Latt\`es-type protocol, the associated rational map on the Riemann (Bloch) sphere, and recall the key characteristics of ergodicity and mixing for pure initial states. Section~\ref{Mixed-state dynamics in a Latt\`es-type ergodic quantum protocol} analyzes the induced nonlinear map for mixed states, identifies the maximally mixed state as an attractor, and characterizes transient purification and angular spreading under preparation noise. Section~\ref{A family of ergodic, non-Latt\`es-type protocols} introduces the one-parameter family $f_p$ and identifies an extended parameter region with ergodic-like pure-state behavior. Section~\ref{Purification and quasi-ergodicity in a non-Latt\`es-type protocol} examines representative non-Latt\`es parameters and shows that genuine purification of mixed states can coexist with rapid angular mixing. We conclude in Section~\ref{Conclusion}.

\section{Pure state dynamics in a Latt\`es-type ergodic quantum protocol}
\label{Pure state dynamics in a Latt\`es-type ergodic quantum protocol}

Iterated quantum protocols that combine entangling unitaries with projective measurement and post-selection induce an inherently nonlinear evolution on the state of a single qubit, or equivalently on an ensemble of uniformly prepared qubits ~\cite{aaronson2005quantum, terno1999nonlinear}. We will refer to the entangling unitary operator as the core unitary determining the properties of a protocol, while the measurements are fixed von Neumann projections in the computational basis and the post-selection criterion is the zero measurement result. The surface of the Bloch sphere represents the full set of pure single-qubit states and can be naturally identified with the Riemann sphere. Consequently, such protocols can always be modeled as iterated complex maps. In particular, any protocol of this kind can be described by a rational map acting on the Riemann sphere, whose iterates determine the evolution of the ensemble. This viewpoint has proved powerful both conceptually and experimentally: measurement-induced nonlinear maps can display exponential sensitivity to initial conditions for all pure inputs, can be engineered to realize broad families of rational maps, including classical examples from complex dynamics ~\cite{Gilyen2016, kalman2018sensitivity}, and have already been implemented in linear-optical settings. ~\cite{zhu2019experimental, qu2021observation}

A particularly distinctive case is the map
\begin{equation}
f(z) = \frac{z^{2} + i}{1 + i z^{2}},
\label{eq:Lattes}
\end{equation}
which arises from the iteration of a non-linear non-unitary quantum operation, which consists of applying a CNOT gate to two identically prepared qubits, creating an entangled pair. One of the qubits is then measured, and the evolution is post-selected on the measurement outcome $0$, introducing a nonlinear update to the remaining qubit~\cite{BechmannP, PORTIK2022127999}.
Finally, the single-qubit unitary
\begin{equation}
U_{L} = \frac{1}{\sqrt{2}}
\begin{pmatrix}
1 & i \\
i & 1
\end{pmatrix}
\end{equation}
is applied to the unmeasured qubit. The correspondence between the map $f$ and the described protocol can be seen by applying a stereographic projection of the normalized pure state onto the complex plane, where the qubit state can be described by a single complex number $z\in\mathbb{C}^{\infty}$, as $\ket{\psi}=N(\lvert 0 \rangle + z \lvert 1 \rangle)$. After the application of the operations in the protocol, the transformed state can be described by a new complex number, given as $f(z)$ of Eq.~(\ref{eq:Lattes}) \cite{KalmanKissJex2018}.

In the language of complex dynamics, Equation~(\ref{eq:Lattes}) defines a Latt\`es map. Concretely, start with the complex torus $\,\mathbb{C}/\Lambda\,$, where $\Lambda \subset \mathbb{C}$ is a two-dimensional lattice. The torus endomorphism given by multiplication by $1 - i$ induces a uniformly expanding map there. Projecting this map to the Riemann sphere $\hat{\mathbb{C}}$ using a Weierstrass elliptic function produces the rational map~\eqref{eq:Lattes}. Because the original torus map is linear and expanding, the induced map on $\hat{\mathbb{C}}$ exhibits uniform stretching and folding, and thus has a positive Lyapunov exponent. This accounts for the protocol’s globally chaotic dynamics on pure states and positions the map of Eq.~\eqref{eq:Lattes} as a canonical example linking iterated quantum protocols to classical rational map dynamics.\cite{Gilyen2016, Milnorbook, Milnor2006}

Ergodicity is a foundational concept in dynamical systems, characterizing dynamics that explore the entire accessible state space in a statistically uniform and indecomposable manner. Intuitively, an ergodic system is one in which, given sufficient time, trajectories visit all regions of phase space in proportion to their measure. More rigorously, with respect to an invariant measure, time averages computed along individual trajectories coincide with ensemble, or spatial averages taken over the state space, implying that the system effectively loses memory of its initial conditions. Ergodic dynamics are often accompanied by stronger chaotic properties such as statistical mixing, whereby initially localized distributions are rapidly stretched and redistributed, leading to an increasingly uniform sampling of the state space. This equivalence implies that the invariant distribution of an ergodic map can be reconstructed either from a single sufficiently long trajectory or from an ensemble of initial conditions evolved for a finite number of steps. In the following, we use both approaches to test ergodicity numerically. In quantum physics, ergodicity is typically discussed in the context of quantum chaos, where the corresponding classical system exhibits chaotic dynamics and the energy eigenstates spread uniformly over phase space. In contrast, the unitary evolution of closed quantum systems preserves state overlaps and therefore prevents ergodic behavior in this strict sense. In the nonlinear quantum protocols discussed here, a new interpretation of quantum chaos emerges. By employing ancillary qubits to perform nonlinear operations on a single qubit, the system effectively behaves as an open quantum system. This arrangement allows genuine chaotic dynamics to arise within a quantum framework. Consequently, phenomena usually associated with classical chaotic systems can be observed and studied in the context of quantum bits, providing a clear bridge between classical and quantum notions of ergodicity and chaos. Genuine ergodic behavior, characterized by a decreasing overlap between evolving states, can thus be identified in such nonlinearly evolving quantum systems.

In our study, we focus on the dynamical system corresponding to iterated rational maps, as these functions effectively describe the time evolution of the  quantum state of a qubit under a nonlinear quantum protocol. In the context of complex dynamical systems, we study iterative rational maps $f: \hat{\mathbb{C}} \to \hat{\mathbb{C}}$ acting on the Riemann sphere. As a first example, we examine the map $f(z) = \frac{z^{2} + i}{1 + i z^{2}}$, whose iterative dynamics on the Riemann (or Bloch) sphere provides a direct link between quantum state evolution and complex dynamical behavior. Such complex maps often reveal a clear contrast between stable and chaotic behavior, formalized by the Fatou and Julia sets. The Fatou set consists of points exhibiting regular or eventually stable dynamics, with trajectories converging to an attractor such as a fixed point or a periodic cycle. In contrast, the Julia set represents the region of chaotic or sensitive dynamics, where infinitesimally close initial conditions lead to significantly different outcomes. For most rational maps, the Julia set forms a fractal subset of the complex plane with zero area measure. However, in special cases this division between order and chaos vanishes: the Fatou set becomes empty, meaning that no point possesses a stable long-term orbit and every point behaves chaotically. In such extreme cases, the Julia set coincides with the entire space. A complex map with an empty Fatou set lacks any attracting periodic cycles or steady orbits, so regardless of the starting point, iterations practically never settle down, almost all trajectories exhibit persistent chaotic behavior (except unstable cycles). These are precisely the ergodic complex maps of interest: they display fully chaotic dynamics across the entire space and typically possess an invariant measure that describes how their overall chaotic behavior is statistically distributed. In the context of complex iterated functions, the invariant measure is a mathematical description of how points are distributed when  iterated asymptotically. For rational maps on the Riemann sphere, this is often the so-called maximal entropy measure, which is essentially the probability measure that remains unchanged when the map is applied. It represents the asymptotic statistical distribution toward which the system converges over time. 

\begin{figure}[htbp]
\centering
\includegraphics[width = \linewidth]{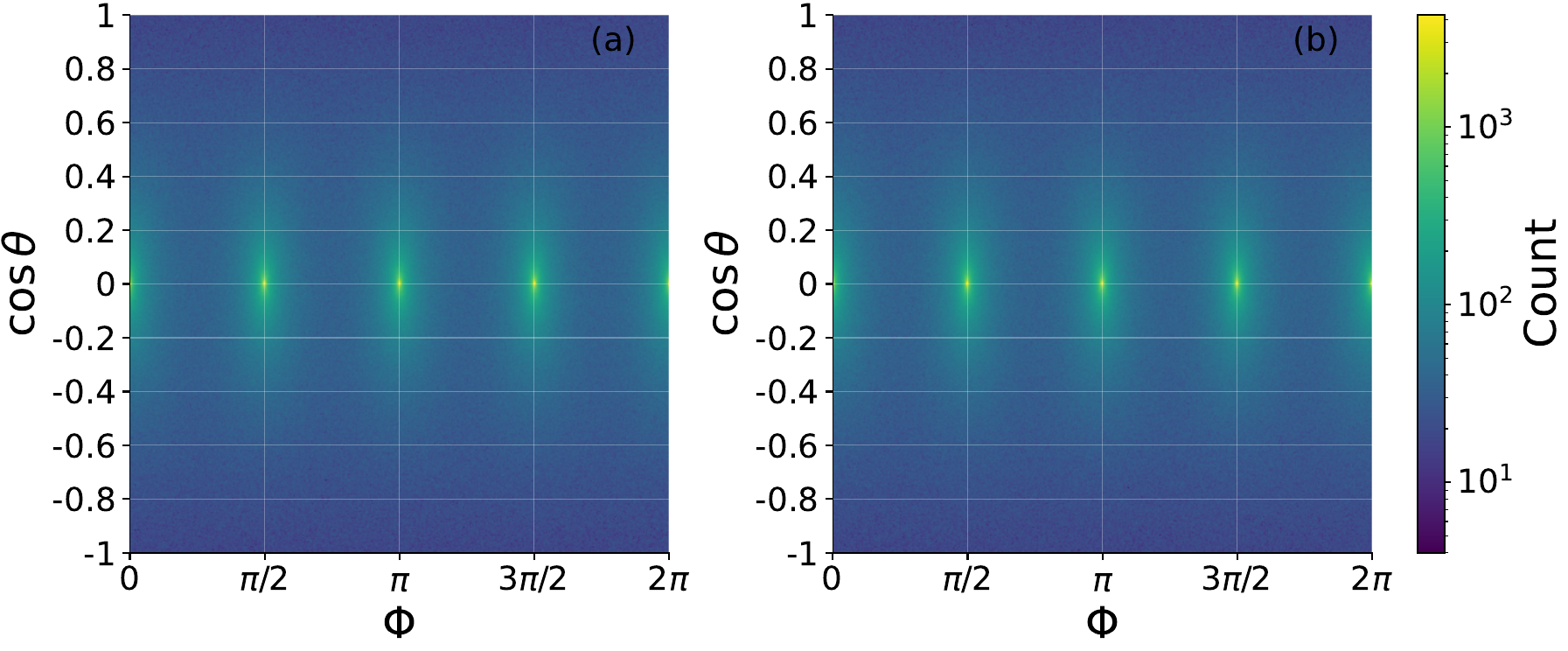}

\caption{Comparison of time-averaged and ensemble-averaged distributions on the Bloch sphere for the case of the Latt\`es map $f(z)$, illustrating the ergodic properties of the iterated map. (a) Time-averaged density: visitation frequency of a forward orbit averaged over 100 randomly chosen initial pure states, each iterated for $10^7$ steps. The Bloch sphere is discretized into a $500 \times 500$ equal-area grid, and the color indicates the average number of visits per cell. (b) Ensemble-averaged density: distribution after $100$ iterations of $10^7$ initial states sampled uniformly from a small region within the Bloch sphere. This process was repeated for 100 randomly chosen regions (of small, equal area), and the resulting histograms were averaged over these trials. Both distributions are visualized using the same grid and color scale.} 
\label{fig:distributions_1j}
\end{figure}

As stated above, the system under consideration exemplifies a special case of an ergodic complex map, namely a Latt\`es map. Latt\`es maps represent a canonical and historically significant class of ergodic complex maps, serving as prototypical examples within the theory of complex dynamics. Their defining property is the complete absence of attracting or neutral cycles; equivalently, their Fatou set is entirely empty. Consequently, the Julia set of such a map coincides with the entire Riemann sphere $\hat{\mathbb{C}}$. This makes Latt\`es maps fully ergodic and strongly mixing examples of complex maps. Under repeated iteration, any small region of points is stretched, folded, and redistributed to cover the entire sphere, resulting in strong statistical mixing. This behavior is illustrated in Fig.~\ref{fig:distributions_1j}, which compares time-averaged distributions obtained from long forward orbits with ensemble-averaged distributions generated by evolving initially localized sets of states. Despite their different constructions, the two distributions are practically indistinguishable within numerical resolution and extend over the full Bloch sphere, illustrating the equivalence of the time- and ensemble-averaged invariant distributions. Remarkably, Latt\`es maps are among the very few rational maps whose natural invariant measure, known as the measure of maximal entropy, is absolutely continuous with respect to the Lebesgue area measure on the sphere. This means that the chaotic orbits of a Latt\`es map are equidistributed according to the uniform area measure, a property not shared by typical rational maps whose Julia sets are fractal and of zero area measure. Consequently, Latt\`es maps occupy a unique position as ergodic systems that realize maximal entropy and metrically uniform chaos on the complex plane. Beyond their mathematical significance, Latt\`es-type dynamics have also been observed in physical systems, particularly in quantum information science. Recent studies have experimantally demonstrated an iterated quantum protocol that is mathematically equivalent to a Latt\`es-type map acting on the Bloch sphere of a qubit. In this photonic realization~\cite{qu2021observation}, an initially localized set of quantum states rapidly spreads across the entire Bloch sphere within only a few iterations, exhibiting deterministic quantum chaos and ergodic evolution.

\section{Mixed-state dynamics in a Latt\`es-type ergodic quantum protocol}
\label{Mixed-state dynamics in a Latt\`es-type ergodic quantum protocol}

In the idealized setting, where the nonlinear quantum protocol can be described exactly by an iterated rational function on the Riemann sphere, the dynamics of pure states exhibit the characteristic statistical properties of ergodic systems. However, from an experimental perspective, the assumption of perfectly pure and identically prepared input states is unrealistic. In real implementations, imperfections in state preparation, decoherence, and other noise cannot be eliminated entirely. Such effects are particularly significant in nonlinear protocols, where even tiny perturbations in the initial state may be exponentially amplified during iteration ~\cite{viennot2020competition}. A physically meaningful way to incorporate preparation noise is to consider mixed initial states described by density matrices rather than pure-state amplitudes. Unlike pure states, which can be fully represented by a single complex coordinate on the Bloch sphere, a mixed qubit state requires three real parameters for its complete description \cite{robustness2023, kalman2018sensitivity}. Accordingly, we consider an ensemble of identically prepared qubits whose initial state is given by the density matrix
\begin{align}
\rho_0 
& =
\frac{1}{\rho_{11} + \rho_{22}}
\begin{pmatrix}
\rho_{11} & \rho_{12} \\
\bar \rho_{12} & \rho_{22}
\end{pmatrix} \\[3mm] \notag
& =
\frac{1}{2}
\begin{pmatrix}
1 + w & u - iv \\
u + iv & 1 - w
\end{pmatrix},
\label{rho_in}
\end{align}
where $u, v, w \in \mathbb{R}$ are the Bloch coordinates, and they satisfy $0 \leq u^2 + v^2 + w^2 \leq 1$. After the application of the CNOT gate and the post-selection step, the remaining qubit evolves according to the nonlinear map
\begin{align}
\rho \rightarrow S(\rho) 
& =
\frac{1}{\rho_{11}^2 + \rho_{22}^2}
\begin{pmatrix}
\rho_{11}^2 & \rho_{12}^2 \\ 
(\bar\rho_{12})^2 & \rho_{22}^2
\end{pmatrix} \\[3mm]
& =
\frac{1}{(1+w)^2 + (1-w)^2}
\begin{pmatrix}
(1+w)^2 & (u - iv)^2 \\ \notag
(u + iv)^2 & (1 - w)^2
\end{pmatrix} ,
\end{align}
where $S(\cdot)$ denotes the nonlinear map induced by the entangling operation and post-selection. If, analogously to the pure-state case, we complete the protocol by applying the single-qubit unitary $U_L$ after post-selection, then the density matrix after one iteration takes the form
\begin{equation}
\rho_1 = U_L\, S(\rho_0)\, U_L^\dagger.
\end{equation}
Because the protocol is iterated, the entire time evolution can again be described in terms of update rules for the Bloch coordinates, where each new state depends on the coordinates of the previous one. Carrying out the transformation explicitly, one finds that the matrix elements of the density operator after the $(k+1)$-th iteration are given by
\begin{align}
\rho_{k+1,\,11} &= \frac{1}{2(1+w_k^{2})}\bigl(1 + w_k^{2} - 2 u_k v_k\bigr), \\
\rho_{k+1,\,22} &= \frac{1}{2(1+w_k^{2})}\bigl(1 + w_k^{2} + 2 u_k v_k\bigr), \\
\rho_{k+1,\,12} &= \frac{1}{2(1+w_k^{2})}\bigl(u_k^{2} - v_k^{2} - 2i w_k\bigr), \\
\rho_{k+1,\,21} &= \frac{1}{2(1+w_k^{2})}\bigl(u_k^{2} - v_k^{2} + 2i w_k\bigr).
\end{align}

The transformed density matrix can again be written in Bloch representation as
\begin{equation}
\rho_{k+1}
=
\frac{1}{2}
\begin{pmatrix}
1 + w_{k+1} & u_{k+1} - i v_{k+1} \\
u_{k+1} + i v_{k+1} & 1 - w_{k+1}
\end{pmatrix},
\end{equation}
where $u_{k+1}, v_{k+1}, w_{k+1}$ now depend on the previous coordinates $u_k, v_k, w_k$ through the nonlinear update rule
\begin{equation}
(u_{k+1}, v_{k+1}, w_{k+1})
=
\left(
\frac{u_k^{2} - v_k^{2}}{1 + w_k^{2}},\;
\frac{2 w_k}{1 + w_k^{2}},\;
\frac{-2 u_k v_k}{1 + w_k^{2}}
\right).
\end{equation}

This three-dimensional nonlinear map governs the evolution of mixed states under the iterated protocol. In the following, we analyze its fixed points, stability properties, and the extent to which ergodic features of the pure-state dynamics persist in the presence of preparation noise.

To understand and assess the properties of the mixed-state evolution, and the behaviour of the corresponding dynamical system embedded in the three-dimensional Bloch sphere, one should examine how the nonlinear protocol affects the purity of the input states. In the literature, there are nonlinear quantum protocols designed specifically for state purification or entanglement distillation. However, for the present protocol we do not expect such behaviour. Indeed, numerical calculations reveal that the only attractive or stable state inside the Bloch sphere is the maximally mixed state. We performed an extensive numerical search for fixed points and periodic orbits of length up to $25$, and found no additional attracting cycles or invariant sets in the interior of the Bloch sphere. This indicates that any mixed initial state eventually (asymptotically) converges to the maximally mixed state. The set of pure states forms an invariant set of the dynamics, meaning that trajectories starting from pure states remain on the surface of the Bloch sphere. Since the maximally mixed state is the only attractive fixed point in the interior, every mixed state is ultimately driven toward the centre of the sphere under repeated iterations.

At the same time, numerical simulations reveal a surprisingly rich behaviour in the evolution of purity. If we consider initial mixed states chosen uniformly from a spherical shell of fixed purity, i.e. random states with the same initial purity, the resulting trajectories do not necessarily converge monotonically toward the maximally mixed state. Instead, a finite fraction of these initial states first become more pure during the early stages of iteration before eventually turning back and converging towards the centre of the Bloch sphere. This transient purification effect appears robust and depends sensitively on the initial purity and the structure of the nonlinear map. In what follows, we characterise this behaviour quantitatively and examine how common such temporarily purifying trajectories are for different initial purities.

\begin{figure}[!htbp]
\centering
\includegraphics[width=0.75\linewidth]{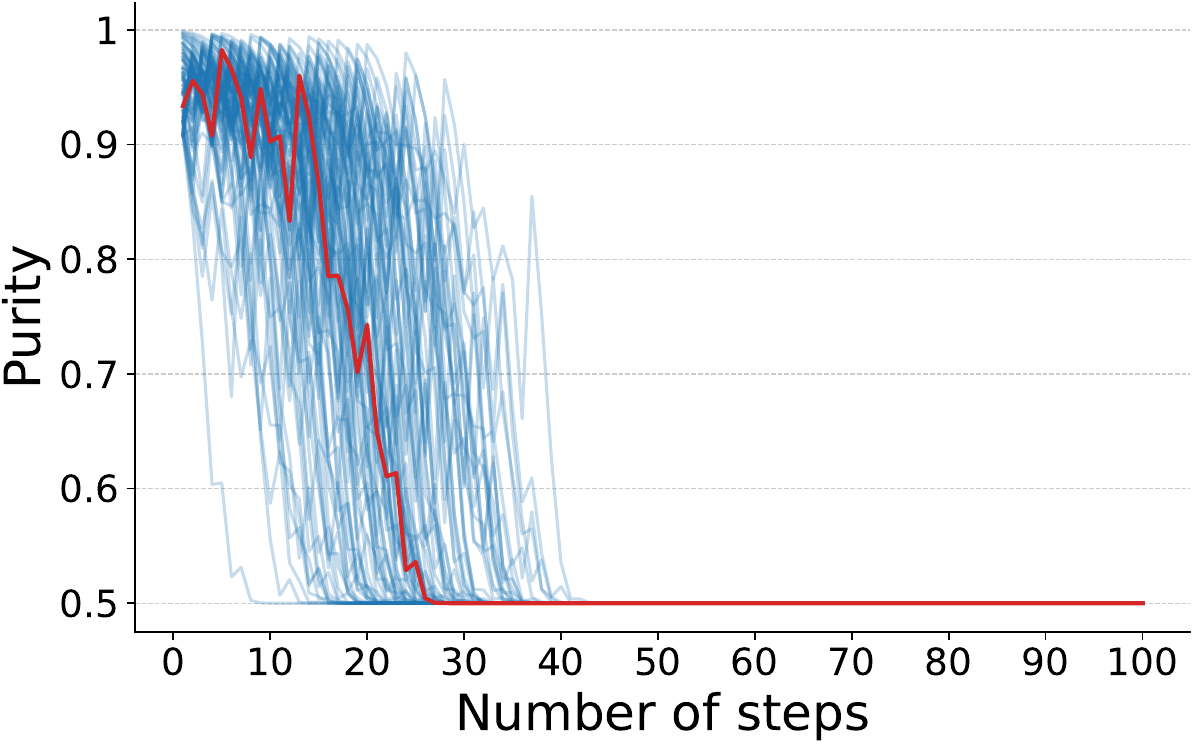}
\caption{Purity of the states along the trajectories of $100$ random initial states sampled uniformly from the spherical surface with initial purity $P_0 = 0.95$. Blue curves show the purity of individual trajectories under the iterated protocol, and the red curve highlights a representative example. The trajectories display a characteristic behaviour, the purity of the state repeatedly increases and decreases over successive iterations, while its overall value gradually declines. Although transient purification occurs, all trajectories ultimately approach the maximally mixed state.}
\label{fig:purity_evol}
\end{figure}

\begin{figure}[!htbp]
\centering
\includegraphics[width=\linewidth]{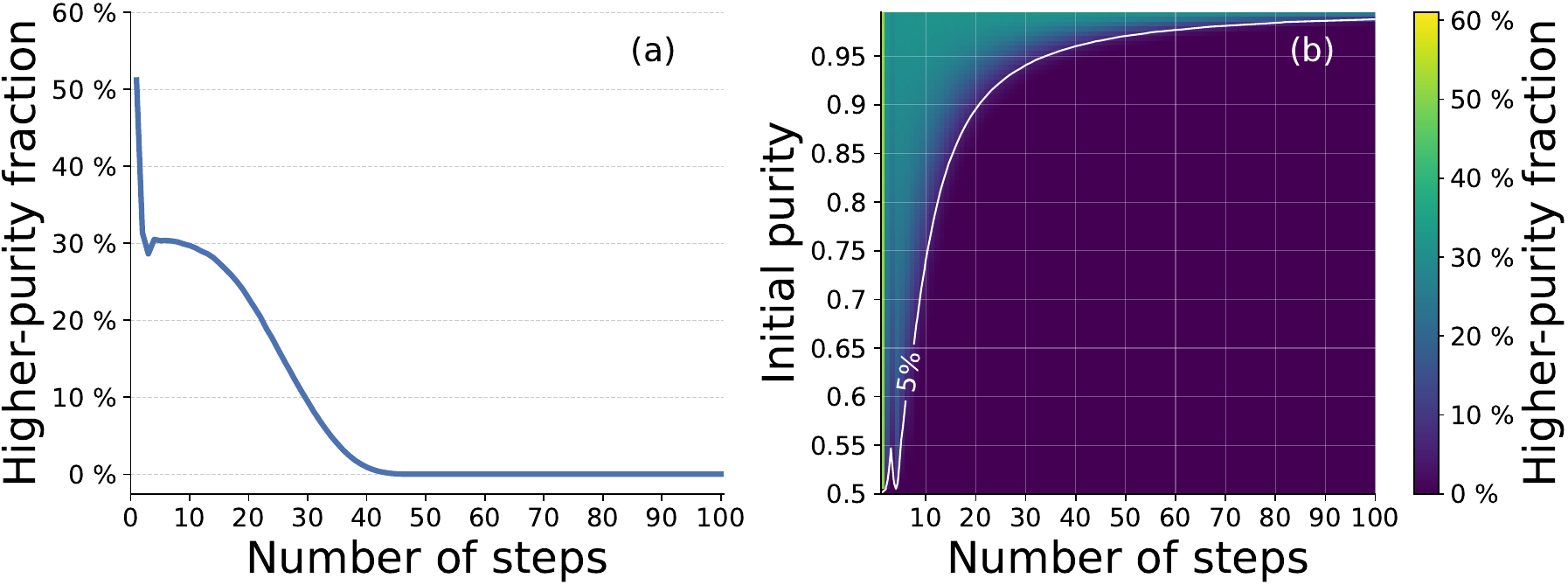}
\caption{Fraction of initial states whose purity increases relative to the previous iteration. Plot (a) shows the fraction for initial purity $P_0 = 0.95$, sampled from $10^6$ random states on the corresponding spherical surface. Plot (b) displays the fraction of states that purify at each step, shown as a function of both the initial purity and the iteration number. Transient purification is most pronounced for states with high initial purity, while it is weak or absent for more mixed initial states. In all cases, the effect disappears as trajectories approach the maximally mixed state.}    
\label{fig:purity_increase_1j}
\end{figure}

The results shown in Figs.~\ref{fig:purity_evol} and ~\ref{fig:purity_increase_1j} illustrate how mixed initial states undergo transient increases in purity. In Fig.~\ref{fig:purity_increase_1j} (a), we fix the initial purity at $P_0 = 0.95$ and uniformly sample $10^6$ random states from the corresponding spherical surface. The curve shows the fraction of these states whose purity becomes higher than it was in the previous iteration. Remarkably, nearly half of the sampled states exhibit an immediate increase in purity during the first iteration, after which this fraction gradually decreases. After approximately $40$–$50$ iterations, none of the trajectories shows any further increase in purity from one step to the next, and all states eventually begin a monotonic approach toward the maximally mixed state. Figure~\ref{fig:purity_increase_1j} (b) provides a broader perspective by scanning over a range of initial purities. For each initial purity $P_0 \in [0.5, 1]$ we determine, at each iteration, the fraction of states whose purity is larger than in the preceding step. The resulting two-dimensional plot reveals a clear pattern: transient purification is most pronounced for states with high initial purity, whereas the effect is weak or entirely absent for more strongly mixed initial states. The bright region around $P_0 \approx 1$ indicates that a significant fraction of nearly pure states become even purer at early iterations before eventually turning around and decaying toward the maximally mixed state. As the number of iterations increases, this bright region rapidly shrinks, demonstrating that the purification effect is short-lived and ultimately suppressed by the global attraction toward the center of the Bloch sphere.

\begin{figure}[!htbp]
\centering
\includegraphics[width = \linewidth]{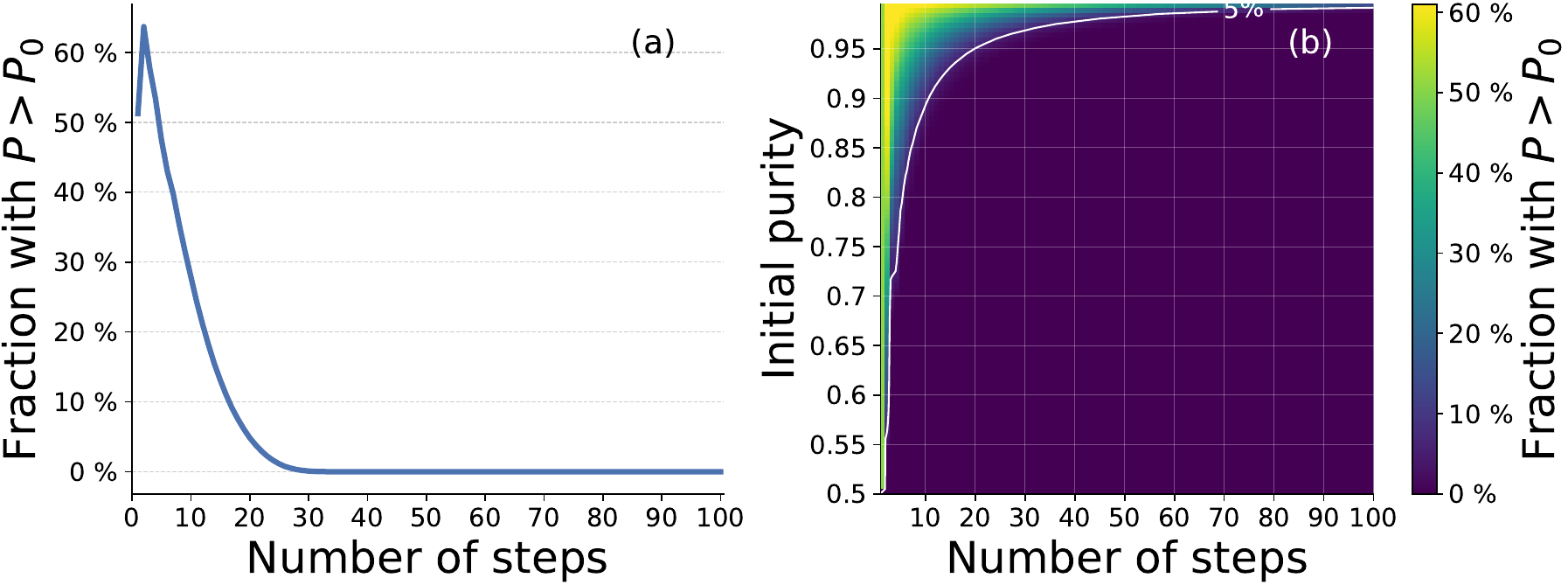}
\caption{Fraction of mixed initial states whose purity becomes higher than their initial value $P_0$ as a function of the iteration number. Panel (a) shows the case $P_0 = 0.95$, panel (b) displays the dependence on both $P_0$ and the iteration number.}
\label{fig:purification_1j}
\end{figure}

The previous analysis showed that mixed states may experience step-to-step increases in purity. However, this does not reveal whether any trajectory ever becomes purer than its own initial purity. To investigate this, we repeated the numerical calculations, but now recorded the fraction of states whose purity exceeds their initial value $P_0$ at a given iteration. The results are shown in Fig.~\ref{fig:purification_1j}. The numerical findings indicate that a significant portion of states (often more than half, in case of moderately mixed initial states) can initially become purer than they started. States with high initial purity are significantly more likely to exceed their starting purity, producing the bright band visible near $P_0 \approx 1$. As the initial purity decreases, this fraction rapidly drops, and for moderately mixed or strongly mixed initial states almost no trajectory ever becomes purer than its starting value after a few iterations. This behavior reflects the fact that for low-purity states, surpassing the initial value requires a substantial increase in the Bloch vector length, which the nonlinear map typically does not generate. Despite these differences, all trajectories exhibit the same qualitative behavior: the fraction of states with $P > P_0$ decreases rapidly with the iteration number and vanishes entirely after approximately $30$–$40$ iterations. Thus, even though many mixed states show a transient purification above their initial purity, this effect is strictly short-lived, and all states eventually converge toward the maximally mixed state.

The pure-state dynamics of the protocol is described by a Latt\`es map, a strongly mixing transformation on the Bloch sphere which causes forward images of any small patch on the sphere to eventually cover the entire surface. In order to examine whether mixed states inherit this characteristic spreading behaviour, we studied how an initially localized set of mixed states expands over the Bloch sphere under iterations. For this, we generated $10^7$ random states inside small rectangular patches whose solid angle corresponds to approximately $1/10\,000$ of the full surface. We selected $100$ such patches at random and we iterated the ensemble of points and monitored their spreading over the Bloch sphere for each of them. Coverage was defined by discretizing the surface into a $500 \times 500$ equal-area grid and determining whether each grid cell had been visited by at least one trajectory in the given step. 

\begin{figure}[!htbp]
\centering

\includegraphics[width=\linewidth]{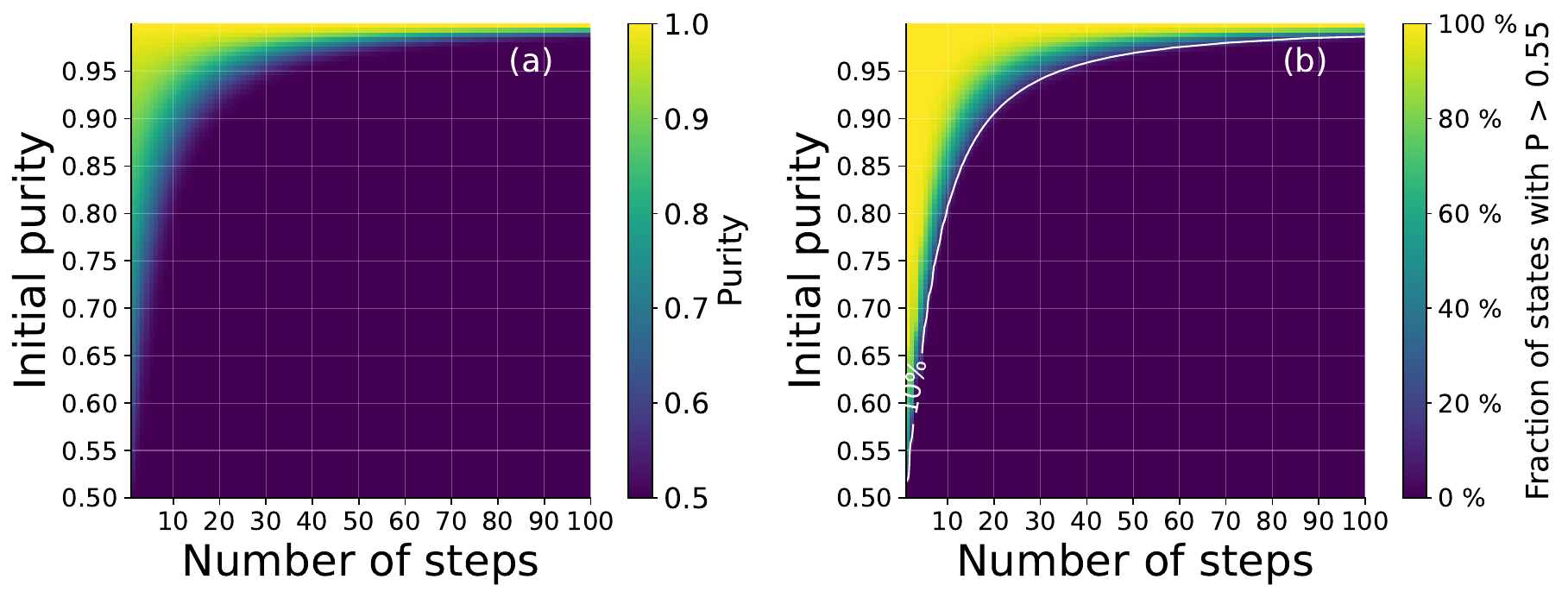}

\caption{(a) Average purity of random initial states under the iterated nonlinear protocol, as a function of the number of steps and the initial purity.
(b) The fraction of states whose purity remains above the purity of the maximally mixed state during the evolution. 
The white contour marks the $10\%$ level.}
\label{fig:average_purity_1j}
\end{figure}

Our simulations show that highly pure initial states retain large average purity for several steps, whereas mixed states rapidly approach the $0.5$ value, corresponding to the maximally mixed state (see Fig. ~\ref{fig:average_purity_1j}). For low initial purities the fraction of states whose purity is larger than $P>0.55$ is large at early steps, but decreases rapidly as all trajectories converge toward the mixed-state attractor. In fact, this fast decay of the purity is the main limitation of this method: After roughly $10$–$20$ iterations, the vast majority of initially mixed states no longer contribute meaningful information about the distribution on the sphere, as they converge toward the maximally mixed state (see Fig. ~\ref{fig:average_purity_1j}). For example, even with initial purity $P_0 = 0.85$, more than $90\%$ of the states reach near-maximal mixing within the first $20$ steps. Lower initial purities suffer even faster contraction. To maintain statistical resolution at later iterations, an exponentially larger number of initial samples would be required.

\begin{figure}[!htbp]
\centering
\includegraphics[width=0.9\linewidth]{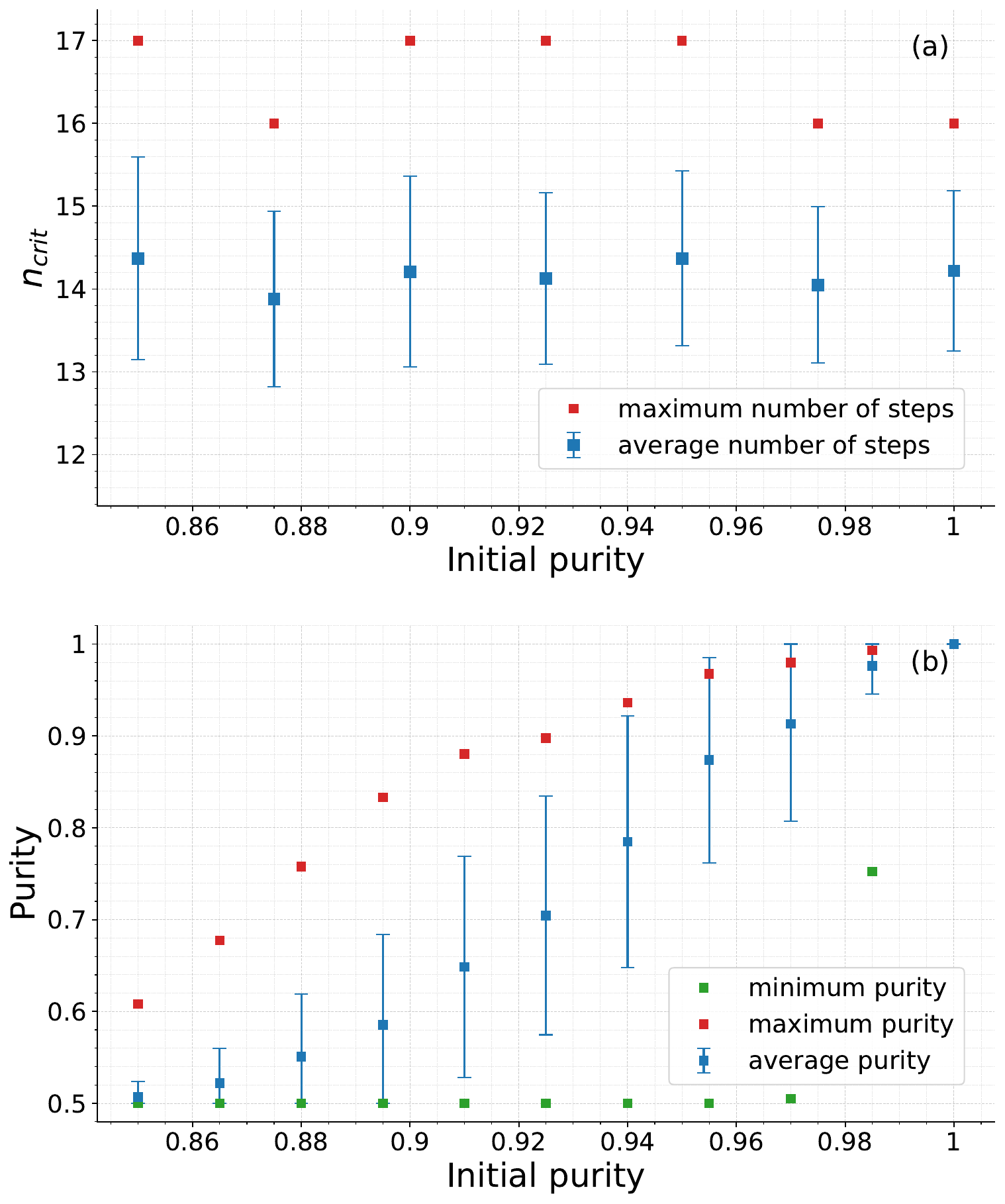}

\caption{(a) Critical number of iterations $n_{crit}$ required for full angular coverage of the Bloch sphere as a function of the initial purity. Blue squares represent the average value of $n_{crit}$ over $100$ randomly chosen initial patches, with error bars indicating the standard deviation. Red squares mark the maximum number of steps needed among all patches for a given initial purity. 
(b) Average purity of the patches after $n_{crit}$ iterations as a function of the initial purity. Blue squares represent the patch-averaged purities, while the error bars show the corresponding standard deviation over the $100$ patches considered in the analysis. Red (green) squares indicate the maximum (minimum) value of the averaged purity over the patches. 
}
\label{fig:n_crit_1j}
\end{figure}

Despite this challenge, our numerical analysis shows that even for such mixed initial states the remaining points continue to spread over the entire surface until they cover the entire solid angle in a finite number of steps $n_{crit}$. Moreover, the value of $n_{crit}$ is essentially independent of the initial purity, as can be seen in Fig. ~\ref{fig:n_crit_1j}~(a). The near-constant behaviour of both the average and the maximum values demonstrates that the spreading time is essentially independent of the initial purity. Fig. ~\ref{fig:n_crit_1j}~(b) shows that the typical purity of the states after $n_{crit}$ iterations already drops close to the maximally mixed value, yet full angular coverage is achieved after a similar number of steps for all initial purities. These results strongly suggest that the spreading behaviour characteristic of the ergodic pure-state dynamics persists, in a practical sense, even when preparation noise is present. Thus, the nonlinear protocol exhibits a form of quasi-ergodicity, namely the ergodic mixing on the surface of the Bloch sphere is robust against preparation noise. While the full mixed-state map acting on the Bloch ball is not ergodic in the same sense as the pure-state map -- since it exhibits a radial drift in purity toward an internal attractor -- the angular component of the dynamics remains strongly mixing. Consequently, ensembles initialized in a small angular patch on a fixed-purity shell spread to cover the entire solid angle corresponding to all Bloch-sphere directions within a finite number of iterations, on a timescale comparable to that of the pure-state dynamics. At the same time, the purity of each initial state in the ensemble gradually, though not monotonically, decreases and approaches the purity of the attractor. In practice, the angular mixing can therefore be meaningfully tracked until the purity becomes sufficiently close to $P = 1/2$, where angular statistics cease to be informative.

\section{A family of ergodic, non-Latt\`es-type protocols}
\label{A family of ergodic, non-Latt\`es-type protocols}

It is a naturally arising question whether, in addition to the Latt\`es-type ergodic protocol discussed above, there exist other nonlinear quantum protocols that also generate ergodic dynamics. A straightforward way to explore this possibility is to study a broader family of protocols that preserve the characteristic structure of the Latt\`es construction. To this end, we introduce a class of nonlinear protocols defined in complete analogy with the original setup: The protocol starts by preparing two identical copies of a single-qubit state and applying a CNOT gate, which serves as the entangling unitary. Next, one of the qubits is measured, and the evolution is post-selected on a chosen measurement outcome (in this case $0$). Similarly to  the Latt\`es protocol, this post-selection produces a simple quadratic update of the complex amplitude describing the remaining qubit. Finally, we apply a general single-qubit unitary transformation to this post-selected state. This last operation introduces additional degrees of freedom into the protocol, enabling a richer variety of nonlinear dynamical behaviours while keeping the overall structure of the protocol similar.

A general single-qubit unitary can be parametrized using three real parameters as
\begin{equation}
U(\xi, \varphi, \omega) =
\begin{pmatrix}
\cos(\xi)\, e^{-i\omega} & \sin(\xi)\, e^{-i\varphi} \\
-\sin(\xi)\, e^{i\varphi} & \cos(\xi)\, e^{i\omega}
\end{pmatrix}.
\end{equation}
However, in the present context the global phase $\omega \in [0,2\pi]$ is physically irrelevant, as its only effect is a uniform redefinition of the computational basis. Therefore, we set $\omega = 0$ without loss of generality. Moreover, instead of using the real parameters $\xi$ and $\varphi$, it is convenient to introduce a single complex parameter $p = \tan(\xi) e^{-i\varphi}$. With this substitution, the unitary transformation takes the compact form
\begin{equation}
U(p) = \frac{1}{\sqrt{1 + |p|^2}}
\begin{pmatrix}
1 & p \\
-\bar{p} & 1
\end{pmatrix},
\end{equation}
where $\bar{p}$ denotes the complex conjugate of $p$. Applying this unitary to the post-selected state yields
\begin{equation}
U_p \left[ N' \left( \ket{0} + z^2 \ket{1} \right) \right]
= N'' \left( \ket{0} + \frac{z^2 - \bar{p}}{1 + p z^2} \ket{1} \right),
\label{eq:Up_action}
\end{equation}
where $N''$ is the appropriate normalization factor. Equation~(\ref{eq:Up_action}) explicitly reveals the rational structure of the induced nonlinear map:
\begin{equation}
\label{eq:fp}
f_p(z) = \frac{z^2 - \bar{p}}{1 + p z^2}.
\end{equation}

Within this framework, the previously analysed Latt\`es-type protocol corresponds exactly to the parameter choice $p=i$. Motivated by this, we now investigate whether nearby parameter values of the form $p=i+\varepsilon$, with $\varepsilon\in\mathbb{C}$, can also give rise to ergodic behaviour. Before turning to numerical results, it is important to note that the family of maps $f_p(z)$ possesses exact symmetries in the complex parameter $p$. These symmetries reveal a deeper conjugacy structure of the entire family. Indeed, the Möbius transformation $h(z)=1/z$ satisfies
\begin{equation}
h\circ f_p = f_{-\bar p}\circ h,
\end{equation}
so the maps $f_p$ and $f_{-\bar p}$ are holomorphically conjugate. Since conjugate rational maps are dynamically equivalent, all quantitive properties of interest, such as the presence or absence of attracting cycles, the structure of the Julia set, the mixing properties, and the Lyapunov exponents, must coincide for parameter values related by this transformation. Consequently, the parameter plane is symmetric under $p\mapsto -\bar p$, i.e., the reflection across the imaginary axis. A second symmetry arises from complex conjugation. For the anti-holomorphic map $h(z)=\bar z$ holds the relation
\begin{equation}
\overline{f_p(z)} = f_{\bar p}(\bar z),
\end{equation}
showing that $f_p$ and $f_{\bar p}$ are anti-holomorphically conjugate. Thus, the same dynamical quantities are invariant under the reflection $p\mapsto \bar p$, corresponding to symmetry across the real axis. 

The same symmetry structure extends also to the mixed-state dynamics generated by the nonlinear protocol. The explicit transformation (update) of the density operator depends on the parameter $p$, this dependence is introduced through the single-qubit unitary $U(p)$. Since each $U(p)$ operator satisfies
\begin{equation}
X U(p) X^{\dagger} = U(-\bar p),
\end{equation}
where $X$ is the Pauli–$X$ operator, conjugating the entire protocol by $X$, therefore maps the mixed-state evolution for parameter $p$ to that for parameter $-\bar p$, establishing the same imaginary-axis reflection symmetry which exists for pure states. Along the same line of thought, since complex conjugation of the unitary gives
\begin{equation}
\overline{U(p)} = U(\bar p),
\end{equation}
and because all other steps of the protocol -- CNOT and projective measurement in the computational basis -- are represented by real matrices, complex conjugation maps the mixed-state update rule with parameter $p$ to the time-evolution corresponding to the parameter $\bar p$. Thus, the mixed-state dynamics are related by a global conjugation symmetry and are therefore invariant under the reflection $p\mapsto \bar p$.

Ergodic rational maps, of which Latt\`es maps are canonical examples, are marked by a collection of distinctive properties, such as complete chaotic coverage of the Bloch sphere, strong sensitivity to initial conditions, rapid mixing, and the absence of attracting periodic cycles or stable regions. While none of these characteristics alone makes a proof of ergodicity, their simultaneous presence provides compelling evidence that the dynamics explores the state space in a statistically uniform and mixing manner. Motivated by this, we numerically investigate the above introduced family of maps $f_p$ for values of $p$ in a neighbourhood of the Latt\`es point ( $p = i$), and examine to what extent the dynamical properties associated with ergodicity persist. Although numerical analysis cannot establish ergodicity in a strict mathematical sense, the collection of results to be presented below strongly suggests that, for a certain range of parameters, the corresponding nonlinear quantum protocols do possess the following four \textit{key qualitative characteristics of ergodic complex dynamics}: (i) they have no attracting cycles, (ii) uniformly sampled points from small patches spread over the entire Bloch sphere (iii) long individual
trajectories densely explore the state space, and (iv) the Lyapunov exponent is strictly positive for all sampled initial conditions. Accordingly, since a rigorous proof of ergodicity for subsets of the analysed family of nonlinear maps is beyond the scope of the present work, we use the term ergodic-like to denote dynamics for which all numerical diagnostics employed in this work are simultaneously satisfied.

As mentioned earlier, a complex rational map that is ergodic on the entire Riemann sphere cannot have any attracting cycles. A standard result in complex dynamics states that the immediate basin of attraction of any attracting periodic point must contain at least one critical point (the point where its derivative is equal to $0$). Therefore, it is sufficient to study the behaviour of the critical orbits in order to detect the presence or absence of attracting cycles. For the family of maps defined by Eq.~\eqref{eq:fp}, it is straightforward to verify that every map has exactly two distinct critical points: $0$ and $\infty$, located at the poles of the Riemann sphere. In order to identify parameter values for which no attracting cycles appear, we numerically generated and analysed the orbits starting from these critical points. For each choice of $p$, we computed $10^7$ iterates of both critical orbits and searched for periodic cycles of length up to $500$. Based on these numerical results, we were able to exclude large subsets of the parameter space where a stable cycle is present, and thus to identify the region in which the map is strongly suggested to have no attracting states.

Beyond the mere absence of attracting cycles, a defining feature of ergodic dynamics is their strong mixing behaviour. Mixing can be examined from two complementary viewpoints, which, by ergodicity, should yield equivalent results. One may analyse the asymptotic visitation statistics of a single long trajectory, or, alternatively, study how a large ensemble of points initially chosen from a small region spreads over the entire state space. Because time averages and ensemble averages coincide for ergodic systems, both approaches should lead to the same conclusion. We applied both methods in order to further narrow the region of parameter values for which the map exhibits ergodic behaviour. To test the time-average approach, we selected $100$ randomly chosen initial pure states and computed, for each, a forward orbit of length $10^7$. For every trajectory we constructed a histogram representing the visitation frequency across the sphere, and then averaged these histograms over the initial states. For the ensemble-based approach, we randomly chose $100$ small patches, each of area equal to $1/10000$ of the full sphere. From each patch we sampled $10^7$ initial states, which then were iterated for $100$ steps. We again constructed histograms to determine how the ensemble spreads across the sphere, and averaged the resulting histograms over all trials. In both analyses, the Bloch sphere was discretized into a $500 \times 500$ equal-area grid.

As a fourth hallmark of ergodic behaviour, we investigated the Lyapunov exponent of the maps $f_p$ as a function of the complex parameter $p$. For a rational map $f_p$, the Lyapunov exponent associated with an initial point $z_0$ is defined by
\begin{equation}
\lambda(z_0) = \lim_{n\to\infty}\, \frac{1}{n}\,
\ln \left| \left. \frac{d}{dz} f_p^{\circ n}(z)\right|_{z=z_0} \right|,
\end{equation}
where $f_p^{\circ n}$ denotes the $n$-fold composition of the map $f_p$ with itself, corresponding to $n$ successive iterations. Using the chain rule, the derivative of the derivate in the expression can be written as
\begin{equation}
\left. \frac{d}{dz} f_p^{\circ n}(z)\right|_{z=z_0}
= \, \prod_{k=0}^{n-1}\frac{2 z_k (1+|p|^2)}{(1+p z_k^2)^2},
\qquad z_{k} = f^{\circ k}_p(z_0).
\label{eq:lyap_chainrule}
\end{equation}

\begin{figure}
\centering
\includegraphics[width=\linewidth]{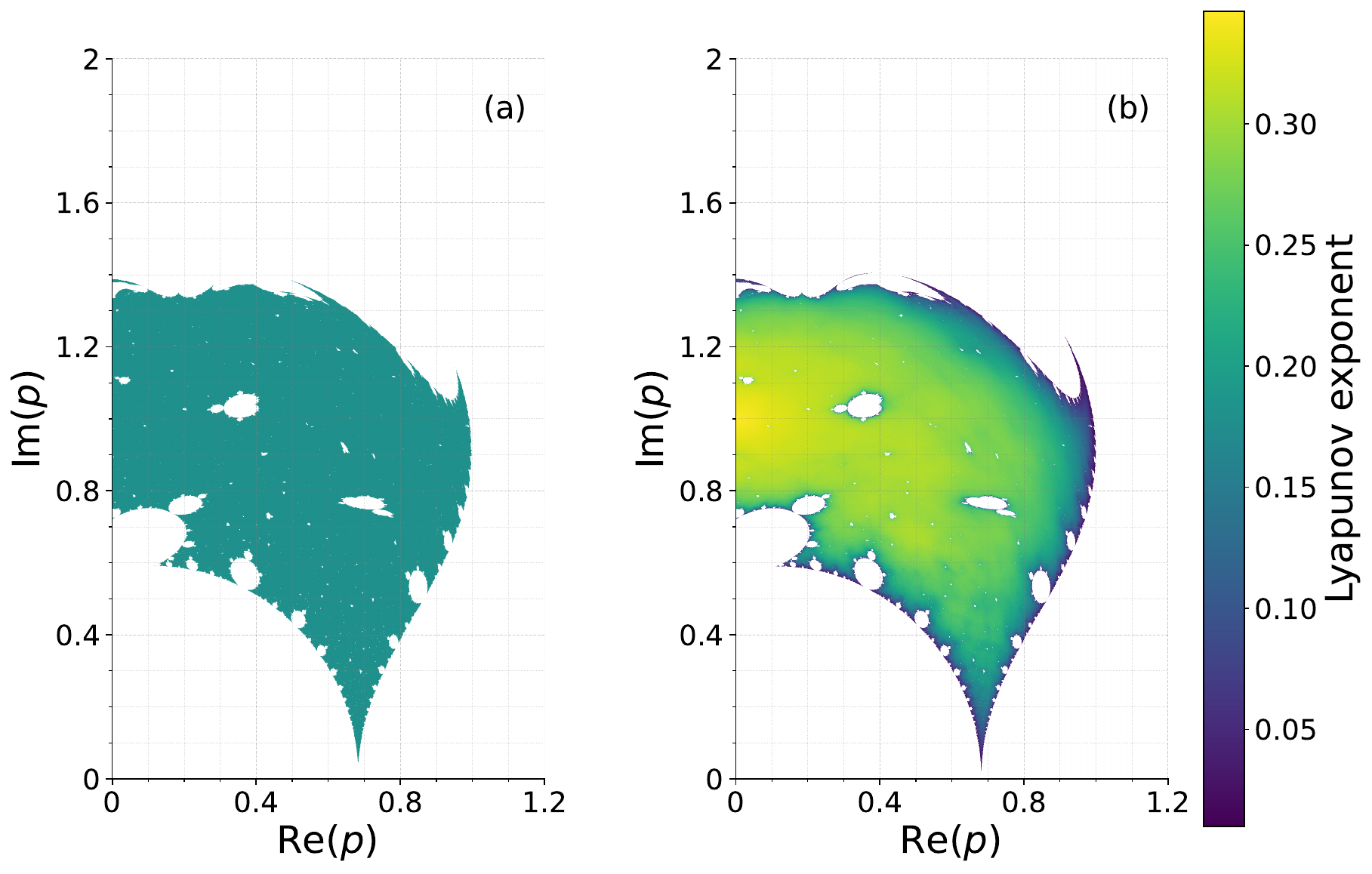}
\caption{(a) Region of the complex parameter space $p \in \mathbb{C}$ for which all four numerical ergodicity criteria are simultaneously satisfied. The highlighted subset corresponds to parameter values for which (i) both critical orbits fail to converge to any attracting cycle, (ii) uniformly sampled points from small patches spread over the entire Bloch sphere, (iii) long individual trajectories densely explore the state space, and (iv) the Lyapunov exponent is strictly positive for all sampled initial conditions. Together, these diagnostics suggest ergodic-like behaviour of the corresponding nonlinear map $f_p(z)$. All four tests were performed on the parameter domain $[0,\,1.2] \times [0,\,2i]$, sampled on a $1200 \times 2000$ grid to resolve fine dynamical structures. The results can be extended to the other quadrants of the parameter space due to the symmetry properties of the maps. Note that the four diagnostics disagree on only $1.414\,\%$ of the sampled points, demonstrating the robustness of the combined ergodicity classification. (b) Numerically estimated Lyapunov exponent of the map $f_p$ as a function of the complex parameter $p$. For each parameter value, the exponent was computed by averaging over $100$ randomly chosen initial states, each iterated for $10^7$ steps. 
} 
\label{fig:ergodicity_lyapunov}
\end{figure}

For maps whose Julia set coincides with the entire Riemann sphere, a situation characteristic of ergodic rational maps we are looking for, the Lyapunov exponent is strictly positive for almost every initial point. Moreover, in the fully ergodic case the exponent typically takes a single well-defined value independent of the choice of initial state, with Latt\`es maps providing the canonical example where $\lambda = \frac{\ln d}{2}$, where $d$ denotes the degree of the rational map, which is $2$ in our case. In order to assess this property, we numerically estimated $\lambda(z_0)$ by iterating the map for orbits of length $10^7$ and accumulating the logarithmic derivative along the trajectory. For each parameter value $p$, we sampled $100$ random initial states uniformly on the sphere and computed the corresponding Lyapunov exponents. We then selected those parameter values for which every sampled trajectory produced a positive Lyapunov exponent, and for which the variation in value across the sample was of the same order of magnitude as the numerical precision (see Fig. ~\ref{fig:ergodicity_lyapunov} (b)). Such behaviour provides strong numerical evidence for a single, parameter-dependent Lyapunov exponent, and thus supports the conclusion that the corresponding nonlinear protocol exhibits ergodic, strongly expanding dynamics.

Figure~\ref{fig:ergodicity_lyapunov}~(a) shows the complex parameters $p$ for which all four numerical criteria introduced above are simultaneously satisfied, indicating ergodic-like behaviour of the corresponding map. These criteria probe complementary aspects of the dynamics, including the absence of attracting cycles, global spreading of initially localized ensembles over the entire Bloch sphere, dense exploration of the state space by long individual trajectories, and exponential sensitivity to initial conditions quantified by a strictly positive Lyapunov exponent. Remarkably, the set of parameters for which all four criteria are fulfilled forms a large, connected region in the complex $p$-plane. This region exhibits an irregular, highly structured boundary, strongly reminiscent of the Mandelbrot set familiar from the theory of iterated quadratic complex functions, reflecting the sensitive dependence of the long-term dynamics on the control parameter.

Within the sampled parameter domain, the four numerical diagnostics show a high level of mutual consistency. For the overwhelming majority of parameter values, either all criteria are satisfied or none of them are. Only a small fraction of points, approximately $1.414\,\%$, lie in regions where at least one criterion fails while the others are fulfilled. These discrepancies occur predominantly near the intricate boundary of the ergodic-like region and are likely attributable to finite iteration times, numerical resolution limits, or the presence of very weakly attracting structures that are difficult to resolve numerically. The strong overlap between the criteria therefore indicates that the identification of the ergodic-like parameter region is robust.

Figure ~\ref{fig:ergodicity_lyapunov}~(b) shows the magnitude of the Lyapunov exponent across the same parameter space. The spatial structure of the Lyapunov map is in good qualitative agreement with the geometric features observed in subplot (a). In particular, regions identified as ergodic-like are characterized by uniformly positive Lyapunov exponents, confirming global exponential instability of trajectories. The exponent reaches its maximal value in the vicinity of the parameter $p=i$, corresponding to the Latt\`es map, in agreement with theoretical expectations. Moving away from $p=i$, the Lyapunov exponent decreases smoothly but remains strictly positive throughout most of the ergodic-like region, indicating that strong chaotic stretching persists even for non-Latt\`es-type protocols.

\section{Purification and quasi-ergodicity in a non-Latt\`es-type protocol}
\label{Purification and quasi-ergodicity in a non-Latt\`es-type protocol}

In the case of the Latt\`es protocol the induced map $f_{p=i}$ is known to be a genuine Latt\`es map, where the ergodic and mixing properties of the dynamics follow from the underlying analytical structure. By contrast, for a generic parameter $p \in \mathbb{C}$ the map $f_p$ is no longer of Latt\`es type, and there is no rigorous theorem guaranteeing ergodicity. Nevertheless, the numerical criteria introduced in the previous section allow us to identify an extended region in parameter space where the dynamics remain strongly chaotic and appear to be ergodic in practice (see Fig.~\ref{fig:ergodicity_lyapunov}~(a)). In order to illustrate how the behavior of the protocol changes as the parameter $p$ is moved away from the Latt\`es point, 
we now focus on the representative choice $p_\ast = 0.4 + 1.2 i $, which lies well inside the numerically determined ergodic region of the parameter space, but is sufficiently far from $p = i$ to exhibit visibly different geometric features.

Let us now analyse in more detail, all four numerical criteria for $p_{\ast}$. First, both critical orbits, starting from $z_0 = 0$ and $z_0 = \infty$ fail to converge to an attracting cycle within $10^7$ iterations. This strongly suggests the absence of attracting cycles and hence that the Fatou set is empty. Second, the Lyapunov exponent is strictly positive for all randomly sampled initial conditions on the Bloch sphere, and the spread of the numerically estimated Lyapunov exponents across different initial conditions is comparable to the numerical uncertainty, indicating that a single, well-defined Lyapunov exponent characterizes the dynamics for almost all initial states. Both properties mirror those of the Latt\`es case.

\begin{figure}[!htbp]
\centering
\includegraphics[width=\linewidth]{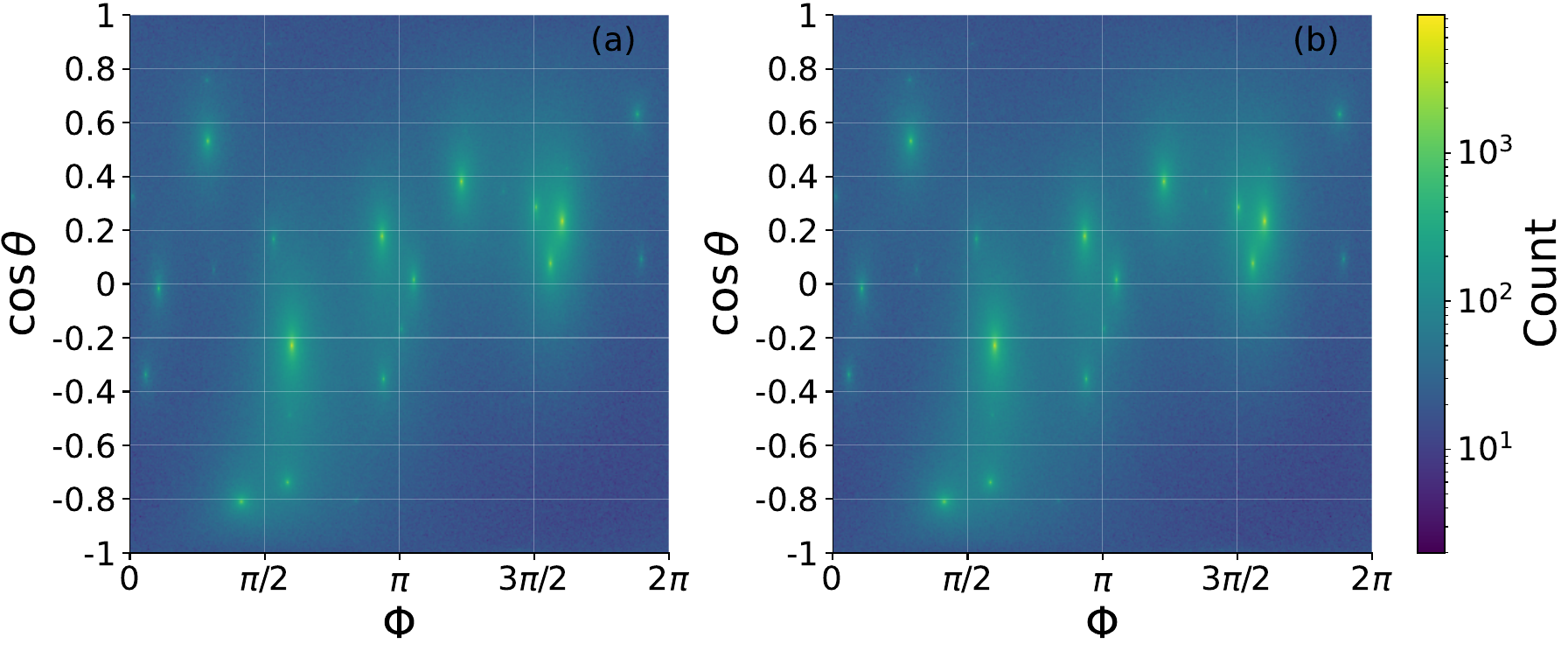}
\caption{Comparison of time-averaged and ensemble-averaged distributions on the Bloch sphere for the case of the protocol corresponding to $p_{\ast}= 0.4+1.2i$. (a) Time-averaged density: Histogram of the visitation frequency of a forward orbit averaged over 100 randomly chosen initial pure states, each iterated for $10^7$ steps. (b) Ensemble-averaged density: Histogram of the distribution after $100$ iterations of $10^7$ initial states sampled uniformly from a small patch within the Bloch sphere, repeated and averaged for 100 randomly chosen equal-area patches. Both distributions are visualized using the same grid ($500 \times 500$), the color indicates the average number of visits per cell. 
}  
\label{fig:distribution_04_12j}
\end{figure}

The strong mixing behavior associated with ergodicity can again be examined from the complementary viewpoints of time-averaged and ensemble-averaged statistics. Figure~\ref{fig:distribution_04_12j}~(a) shows the time-averaged visitation histogram obtained from $100$ randomly chosen initial pure states, each iterated for $10^7$ steps, under the map $f_{p_{\ast}}$. Figure~\ref{fig:distribution_04_12j}~(b) displays the ensemble-averaged distribution produced by iterating $10^7$ initial points sampled from small patches of equal area for $100$ steps. Although the invariant density is visibly anisotropic and structured, in clear contrast to the near-rotational symmetry observed in the Latt\`es case (see Fig. ~\ref{fig:distributions_1j}), the time-averaged and ensemble-averaged histograms remain practically indistinguishable within numerical precision. The fact that both a single long forward trajectory and an initially localized ensemble spread to produce the same invariant distribution, together with the near-identity of the corresponding time and space averages, provides strong numerical evidence that the defining features of ergodic dynamics persist for this parameter value as well.

\begin{figure}[!htbp]
\centering
\includegraphics[width=\linewidth]{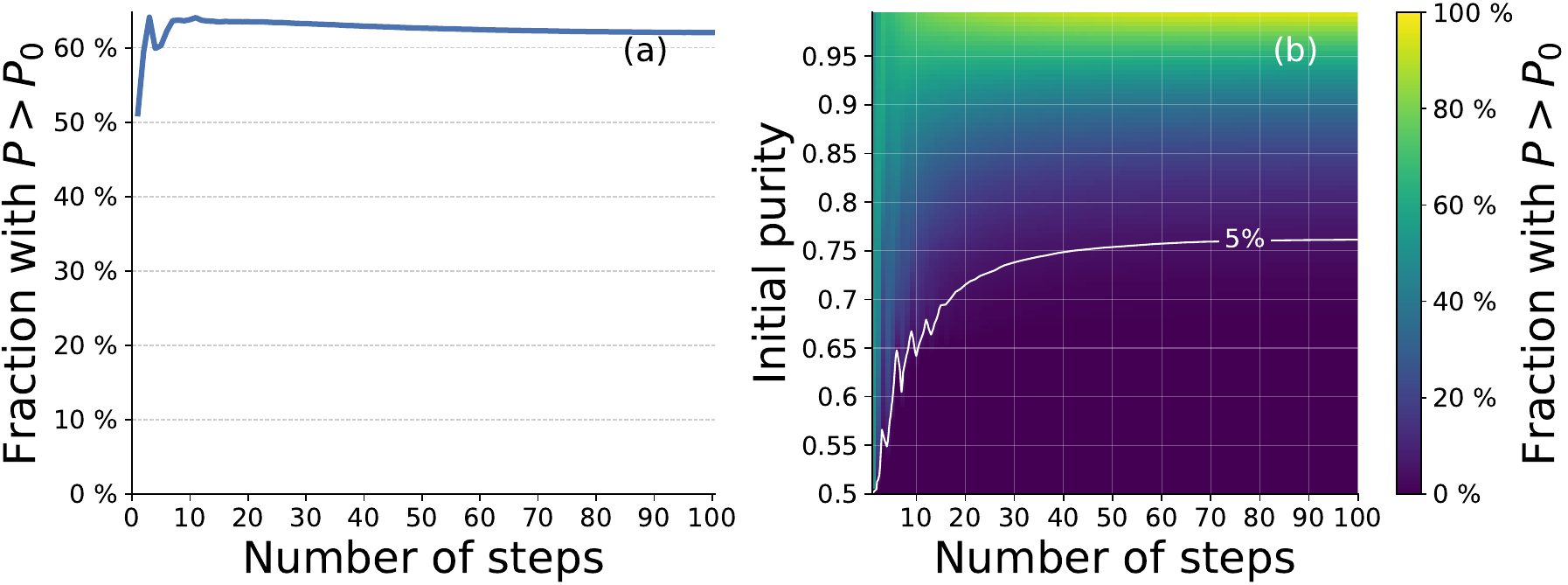}
\caption{Fraction of mixed initial states whose purity becomes higher than their initial value $P_0$ as a function of the iteration number for the protocol corresponding to $p_{\ast} = 0.4 + 1.2i$. (a) The behaviour for $P_0 = 0.95$. 
(b) Dependence on both $P_0$ and the iteration number. 
}
\label{fig:purification_04_12j}
\end{figure}

To assess the full dynamical behaviour of the protocol for the parameter  $p_{\ast}$ we examined how the purity of random mixed initial states evolves under iteration. While the above key features of the pure-state dynamics was found to be analogous to the Latt\`es case, one significant new phenomenon emerged when observing the mixed-state dynamics: A finite fraction of randomly sampled initial mixed states repeatedly purifies and is driven toward the surface of the Bloch sphere, despite the fact that no attracting set or cycle can be identified within the pure states (see Fig.~\ref{fig:purification_04_12j}). As can be seen from Fig.~\ref{fig:purification_04_12j}~(a), after an initial transient, the fraction of states with $P > P_0$ does not decay to zero but instead settles to a finite asymptotic value, indicating that a non-negligible subset of trajectories exceeds its initial purity throughout the evolution. If one examines states for a broader range of initial purities (Fig.~\ref{fig:purification_04_12j}~(b)), it can be seen that for high initial purities this fraction is substantial, but even for moderately mixed initial states $(P_0 \gtrsim 0.7)$ a persistent portion of trajectories surpasses their initial purity. This behaviour demonstrates that, unlike in the Latt\`es case, the nonlinear dynamics for this parameter value attract part of the ensemble back toward the surface of the Bloch sphere, leading to real purification rather than purely transient ones. As a consequence, the completely mixed state is no longer the unique attractor for the interior of the Bloch sphere, the open set of mixed states does not form its basin of attraction. This behaviour represents a qualitative deviation from the Latt\`es case and highlights a distinct structure in the mixed-state dynamics for the shifted $p$ parameters.

\begin{figure}[!htbp]
\centering
\includegraphics[width=\linewidth]{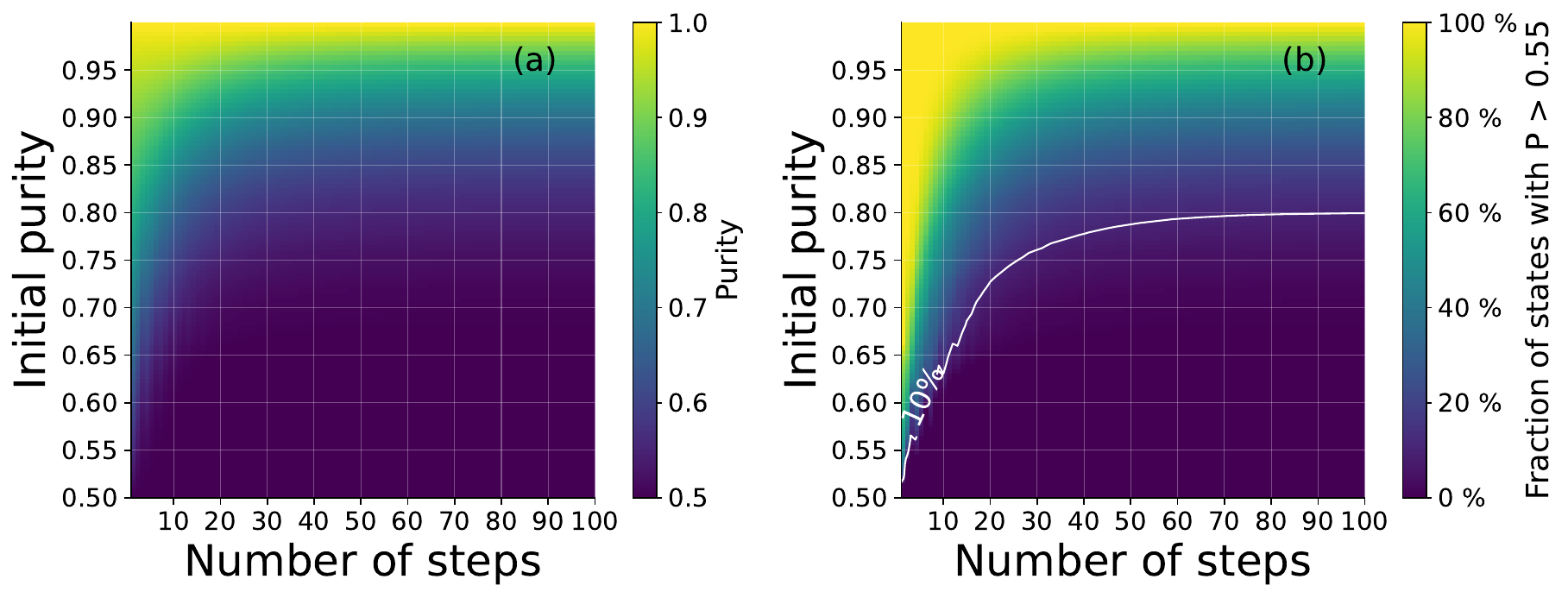}

\caption{Purity dynamics of random initial states under the iterated nonlinear protocol for $p_{\ast} = 0.4 + 1.2i$. (a) Average purity after each iteration. 
(b) Fraction of states with $P>0.5$ ($0.5 + 10 \%$). 
The white contour marks the $10\%$ level, indicating the minimal initial purity needed for a non-negligible fraction of states to retain $P>0.5$.}
\label{fig:avergae_pruity_04_12i}
\end{figure}

In order to further characterise the mixed-state dynamics of the protocol, we examined how the average purity of random initial states evolves under the protocol. As shown in Fig. ~\ref{fig:avergae_pruity_04_12i}~(a), the average purity behaves qualitatively differently from the $p=i$ case: It no longer decays uniformly toward the maximally mixed state. States that begin with high purity remain close to the surface for many steps and may even reach it within numerical precision, while for moderately mixed initial purities a finite subset of trajectories is likewise drawn toward the surface, even though the majority is rapidly attracted to the maximally mixed state. Fig. ~\ref{fig:avergae_pruity_04_12i}~(b) further illustrates this behaviour by displaying the fraction of states whose purity stays above $P=0.5$. This fraction quickly vanishes for low initial purities but remains substantial for sufficiently pure initial states, revealing a persistent set of trajectories that converge toward the surface.

\begin{figure}[!htbp]
\centering
\includegraphics[width = 0.9\linewidth]{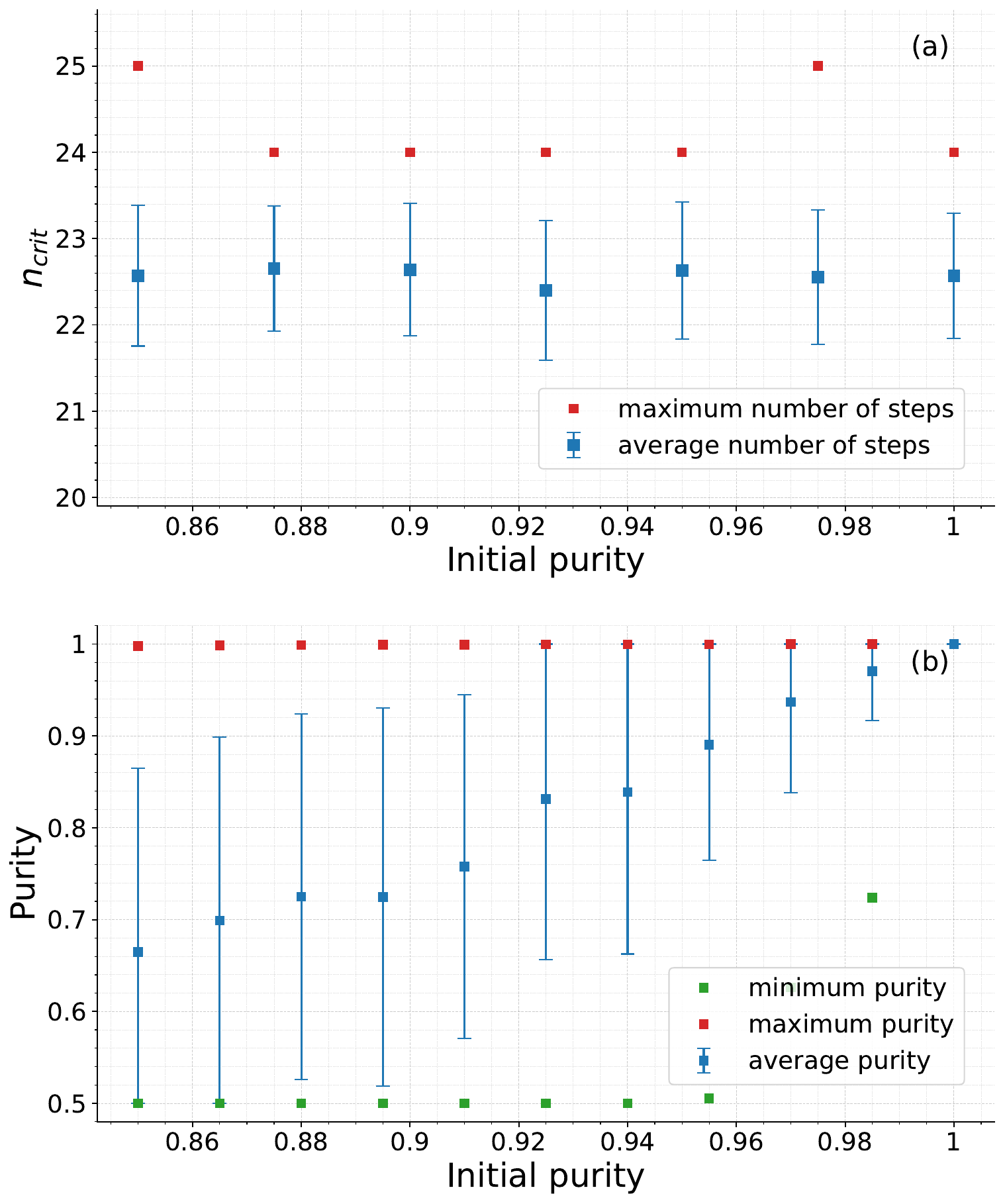}

\caption{(a) Critical number of iterations $n_{crit}$ required for full angular coverage of the Bloch sphere as a function of the initial purity. Average (maximum) values are marked by blue (red) squares. Error bars indicate the standard deviation of the values. 
(b) Average purity of the ensembles after $n_{crit}$ iterations as a function of the initial purity. Minimum, average, and maximum purities are denoted by green, blue and red squares, respectively.  
}
\label{fig:n_crit_04_12j}
\end{figure}

We also analysed the spreading behaviour for the perturbed parameter value $p_{\ast}$ and found that although the mixed-state dynamics differ in important details, the ensemble still exhibits rapid angular spreading, and the characteristic number of steps required for the iterates to cover the full solid angle is again largely independent of the initial purity, similarly to the Latt\`es case. The critical number of steps required to spreading is slightly larger for $p_{\ast}$ than for $p=i$, in agreement with our findings for the pure-state dynamics, where the same trend was observed (see Fig. ~\ref{fig:n_crit_04_12j} (a)). These results demonstrate that effective mixing on the Bloch sphere persists across parameter values and is not restricted to the Latt\`es map. In this sense, the nonlinear protocol continues to display quasi-ergodic behaviour even away from the original parameter choice. 
Observing the findings presented on Fig.~\ref{fig:n_crit_04_12j}~(b), one can see that, in contrast to the $p=i$ case, where the ensemble-averaged purity always remained strictly below unity, the new parameter choice results in a qualitatively different behaviour: For sufficiently high initial purities, the maximal ensemble purity reaches $P=1$, demonstrating that an entire subset of trajectories is now driven all the way to the surface of the Bloch sphere, even for an initial purity as low as $P_0 = 0.85$, similarly to what we have seen in Fig~\ref{fig:avergae_pruity_04_12i}. 

\begin{figure}[!htbp]
\centering
\includegraphics[width=0.6\linewidth]{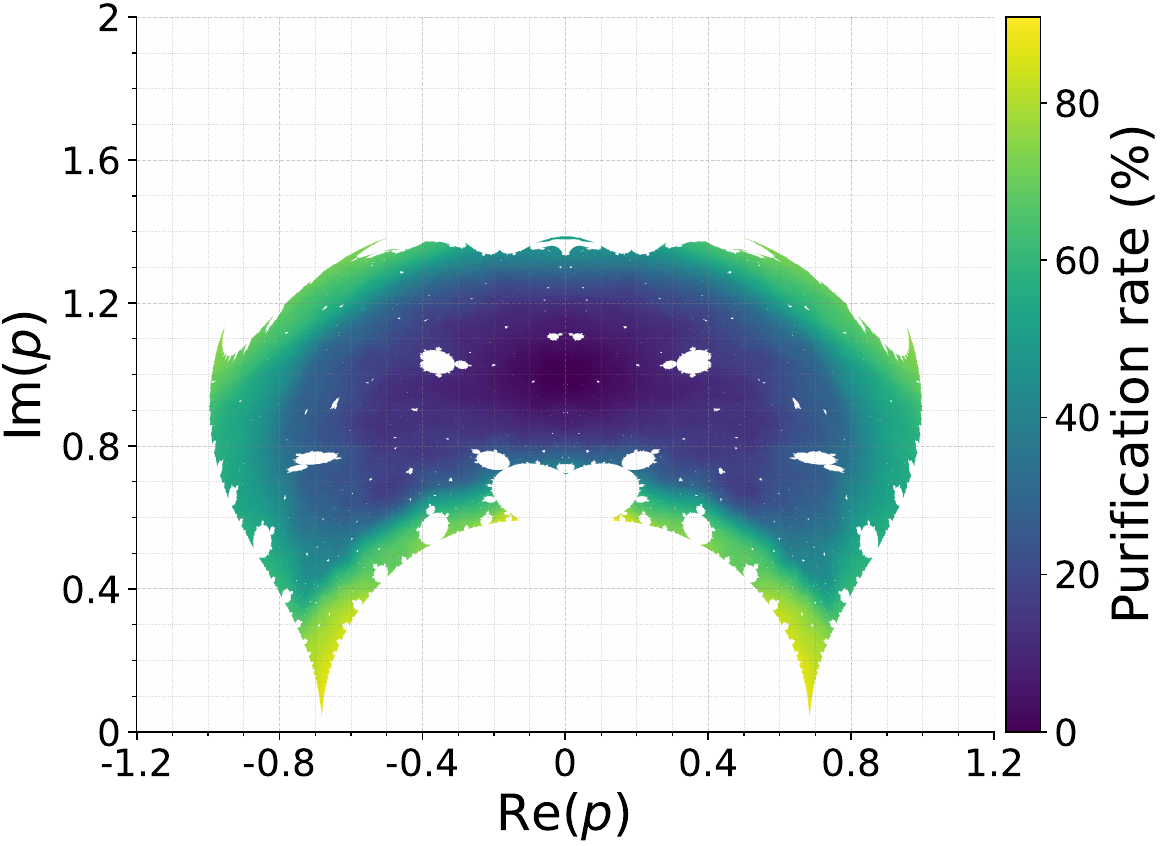}
\caption{Purification rate as a function of the complex parameter $p$. The colour scale indicates the fraction of randomly sampled mixed initial states that converge to the surface of the Bloch sphere under iteration. 
The numerical computation was carried out over the parameter domain $[-1.2, 1.2] \times [0, 2i]$, sampled on a $2400 \times 2000$ grid to ensure sufficient resolution for detecting fine structural features in the dynamics. For each parameter value, we uniformly sampled $10^7$ initial states, which were then iterated $10\,000$ times or until all states had converged.}    
\label{fig:purification_rate}
\end{figure}

The numerical findings presented above show that the parameter values for which all four ergodicity indicators are satisfied exhibit dynamics broadly similar to the original Latt\`es protocol corresponding to $p = i$. The main difference is the emergence of mixed states that purify under the iteration of the protocol. For the example, in the $p_{\ast} = i + (0.4 + 0.2i)$ case, we observed that a finite fraction of initially mixed states converges to the surface of the Bloch sphere and never starts approaching the maximally mixed sate again. This observation raises the question whether there exist parameter values that behave ergodically on the space of pure states yet display fundamentally different dynamics for mixed inputs. To address this possibility, we examined the long-term evolution of $10^7$ randomly chosen mixed states for every parameter value in the region where all four ergodicity criteria were met (see Fig.~\ref{fig:purification_rate}). Remarkably, for the vast majority of these parameters, a finite subset of trajectories converged to the surface, indicating complete purification. There was only a very small fraction of parameters (less than $1\%$), located in the immediate vicinity of $p = i$ (and  $p = -i$), where we observed no purifying trajectories, although this may have been due to numerical precision limitations. Our results also suggest that these ergodic-like protocols inherit a key structural characteristics of the Latt\`es case, namely that they possess no attractive fixed points or cycles of intermediate purity $0 < P < 1$.

\section{Conclusion}
\label{Conclusion}

In summary, we have analyzed the evolution of a family of iterated, nonlinear quantum protocols acting on an ensemble of single qubits with ergodic and quasi-ergodic evolution. The core unitary of the protocol consists of a CNOT gate and a parametrized single qubit gate. For pure inputs, a specific choice of the parameter ($p=i$) realizes the Latt\`es map $f(z)=(z^2+i)/(1+iz^2)$, providing an example where the Julia set is the full Riemann (Bloch) sphere and the dynamics is uniformly expanding and mixing. We have presented an extensive numerical analysis to confirm that the corresponding dynamical system is ergodic, meaning that a localized ensemble spreads to cover the whole state space after a finite number of steps, and long forward orbits reproduce the same invariant distribution as ensemble averages, consistent with the equivalence of time and space averages. 

Extending the analysis to the full family of maps $f_p(z)=(z^2-\bar p)/(1+p z^2)$ induced by varying the final single-qubit unitary utilized in the protocol, we identified an extended region in the complex $p$-plane where our numerical findings about four independent characteristics of the map (namely nonconvergence of critical orbits to attracting cycles, rapid angular spreading of small ensembles, agreement of time- and ensemble-averaged distributions, and uniformly positive Lyapunov exponent) indicate ergodic-like behavior on pure states. The strong agreement among these criteria suggests that ergodic dynamics is not confined to the Latt\`es point but persists across a sizeable parameter regime of experimentally accessible protocols.

We found that for mixed initial states, the Latt\`es case exhibits a qualitatively different structure compared to the pure-state dynamics: The maximally mixed state becomes an attractive fixed point for the interior of the Bloch sphere, and no additional attracting cycles up to period $25$ was found. Consequently, all mixed states converge asymptotically toward the maximally mixed state.  Our numerical simulations have revealed that the convergence is not strictly monotone in the purity parameter, a sizeable fraction of initially high-purity states shows transient purification before ultimately converging to the fully mixed attractor. Despite this contraction in the radial (purity) direction, the angular component of the dynamics retains the rapid spreading characteristic of the ergodic pure-state map, motivating a practically relevant notion of quasi-ergodicity that remains robust under a small amount of noise. 

Moving away from $p=i$ while staying in the numerically ergodic region, we found that for typical ergodic-like parameters, a non-negligible subset of mixed initial states is driven toward the surface of the Bloch sphere and eventually get purified, thus the maximally mixed state is no longer the only attractive state in the dynamics. This coexistence of ergodic-like pure-state dynamics with persistent purification for mixed inputs highlights an unexpected interplay between two distinct characteristics of these quantum protocols, namely the chaotic mixing on the sphere and purification for mixed initial states.

We note that in a recent proposal for an effective Bell pair distillation scheme~\cite{Kalman2025}, a similar type of core unitary is acting locally, in both distant laboratories. A few iterations of that scheme leads to efficient distillation of a specific Bell pair. The distillation scheme can be extended to three-qubit GHZ states as well~\cite{Rozgonyi1,Rozgonyi2}. In contrast to the presented one qubit scheme, leading to chaotic dynamics, the distillation scheme is a two- (or more-) qubit scheme with fast convergence to the desired state. The fact that both schemes are based on the same type of core unitary (a CNOT combined with a single qubit gate) underlines the importance of understanding the iterative dynamics for various parameters of the core unitary. 

Looking ahead, it would be valuable to further study the mixed-state dynamics beyond the Latt\`es map, in particular, to characterize the purification induced by these ergodic protocols. A natural next step could be to extend the present approach to higher-order single-qubit protocols (processing more copies per round), which could reveal whether ergodic-like behavior and mixed-state purification persist, compete, or disappear with increasing nonlinearity. Better insight into these processes may open routes to applications such as noise-robust state preparation, benchmarking, or controlled generation of pseudo-random single-qubit ensembles.

\begin{acknowledgments}
We acknowledge support from the National Research, Development and Innovation Office of Hungary, project No. TKP-2021-NVA-04 and the Quantum Information National Laboratory of Hungary (Grant No. 2022-2.1.1-NL-2022-00004). TK is grateful for the support from the ‘Frontline’ Research Excellence Programme of the NKFIH (Grant No. KKP133827). AP also acknowledges support from the National Research, Development and Innovation Office of Hungary, Project No.\ C2245617 (KDP-2023).
\end{acknowledgments}

\section*{References}
\bibliography{references}

\end{document}